# Anomalous Hall Effect in Thin Film of the Weyl Antiferromagnet Mn$_3$Sn

Tomoya Higo [1,2], Danru Qu [1], Yufan Li [3], C. L. Chien [3], Yoshichika Otani [1,2,4], and Satoru Nakatsuji [1,2,5*]

[1] Institute for Solid State Physics, University of Tokyo, Kashiwa, Chiba 277-8581, Japan

[2] CREST, Japan Science and Technology Agency (JST), 4-1-8 Honcho Kawaguchi, Saitama 332-0012, Japan

[3] Department of Physics and Astronomy, Johns Hopkins University, Baltimore, Maryland 21218, USA

[4] Center for Emergent Matter Science, RIKEN, Wako, Saitama 351-0198, Japan

[5] Institute for Quantum Matter and Department of Physics and Astronomy, Johns Hopkins University, Baltimore, Maryland 21218, USA

E-mail: satoru@issp.u-tokyo.ac.jp

**Abstract**

The Weyl antiferromagnet Mn$_3$Sn has recently attracted significant attention as it exhibits various useful functions such as large anomalous Hall effect that are normally absent in antiferromagnets. Here we report the thin film fabrication of the single phase of Mn$_3$Sn and the observation of the large anomalous Hall effect at room temperature despite its vanishingly small magnetization. Our work on the high-quality thin film growth of the Weyl antiferromagnet paves the path for developing the antiferromagnetic spintronics.

Recent rapid growth in the information technology has demanded the spintronic devices to acquire a higher integration density and faster data processing, for instance, in non-volatile magnetic memory devices [1]. For these requirements, antiferromagnets have attracted much attention recently as an important material for next generation memory devices, because they generate no perturbing stray field and have much faster spin dynamics than ferromagnets, [2–14]. On the other hand, it is also true that the absence of magnetization, which is the reason for the above benefits, have made it notoriously difficult to control and otherwise produce only very weak responses to detect them.

In this context, the recent discovery of functional antiferromagnets $D0_{19}$ Mn$_3X$ ($X$ = Sn, Ge), the first example of a topological "Weyl magnet", has attracted a lot of attention [8-10,15-19]. This is because it exhibits a variety of useful functions that have never been seen in antiferromagnets at zero magnetic field, which include (1) anomalous Hall effect [8-10] (2) anomalous Nernst effect [15,16] and (3) magneto-optical Kerr effect [18] at room temperature. Moreover, they are controllable by magnetic field and thus can be used for designing antiferromagnetic spintronics and energy harvesting technology. It has been shown that the antiferromagnet exhibiting these effects hosts the topological state called magnetic Weyl metal, characterized by relativistic Weyl fermions [17,19]. The discovery of the Weyl metal state in an antiferromagnet represents the opening of another chapter of applied research using the functional antiferromagnets and it is thus highly important to fabricate a high-quality thin film of the Weyl antiferromagnet.

$D0_{19}$ Mn$_3$Sn has a hexagonal Ni$_3$Sn-type structure with space group $P6_3/mmc$ (Fig. 1(a)). The basal plane projection of each (0001) plane consists of a kagome lattice of Mn magnetic atoms, and the associated geometrical frustration manifests itself as an inverse triangular spin structure, which is antiferromagnetic ordering with very small net magnetic moments of ~3 m$\mu_B$/Mn, below the Néel point $T_N$ ~ 420 K [20,21]. As shown in Fig.1(b), this unique antiferromagnetic spin structure within the (0001) plane can be characterized by a ferroic ordering of a cluster-octupole moment consisting of the non-collinear spins on the Kagome bilayer [22]. Thus, unlike a normal antiferromagnetic state, this breaks time-reversal symmetry (TRS) macroscopically. In ferromagnets, it is the spontaneous magnetization that breaks TRS and the anomalous Hall effect appears proportionally to magnetization based on the mechanisms including side jump scattering, skew scattering, and intrinsic (Berry curvature) effects [23]. In Weyl magnets, on the other hand, the anomalous Hall effect arises from the intrinsic mechanism [8-10,15-19]. In $Mn_3Sn$, the TRS breaking by the non-collinear antiferromagnetic order leads to the formation of Weyl points [15-19] and the sizable Berry curvature or fictitious field corresponding to the external magnetic field of ~100 T in the momentum space even with a vanishingly small FM moment of ~3 m$\mu_B$/Mn [17,19,22]. As a result, this magnetic Weyl metal state exhibits the large topological responses (e.g. the anomalous Hall effect), comparable to those in ferromagnets.

Here we report the fabrication of a high-quality thin film of $Mn_3Sn$ that exhibits the anomalous Hall effect at room temperature. We find that a single phase polycrystalline thin film of $Mn_3Sn$ can be deposited onto Si/SiO$_2$ substrate. Our measurements of the Hall effect indicate that the film shows approximately half size of the anomalous Hall effect compared to the values reported for single crystals [8,15-18]. Our work on the high-quality thin film growth paves the path for developing antiferromagnetic spintronics and energy harvesting technology by using Weyl magnets.

For the thin film fabrication, we employ DC magnetron sputtering. The films are deposited at room temperature onto Si/SiO$_2$ substrate from $Mn_{2.5}Sn$ target in a chamber with a base pressure of $5 \times 10^{-7}$ Pa. The sputtering power and Ar gas pressure are 60 W and 0.3 Pa, respectively. For this condition, our X-ray reflectivity measurements find the deposition rate to be ~0.2 nm/s. During the deposition, we rotate the sample holder along its normal axis to ensure a uniform composition and thickness. After deposition at room temperature, we anneal the film at 500 °C for 1 h. The annealing procedure after the deposition crystallizes the as-deposited amorphous film into a polycrystalline form of the $Mn_3Sn$ film. The composition of the film is estimated to be $Mn_{3.05}Sn_{0.95}$ by using the inductively coupled plasma atomic emission spectroscopy (ICP-AES). We note that the Mn composition of our film is within the range of the composition,

Mn$_{3.02}$Sn$_{0.98}$-Mn$_{3.15}$Sn$_{0.85}$, where the D0$_{19}$ Mn$_3$Sn phase is reported to be stable due to the presence of excess Mn, which randomly occupies the Sn site [8,20,21]. Structural analysis is performed by using the X-ray diffraction (XRD) spectra. The magnetization is measured with a commercial SQUID magnetometer (MPMS, Quantum Design). To estimate the signal from the samples, the diamagnetic contribution of the Si/SiO$_2$ substrate is separately measured and subtracted from the magnetization. Both longitudinal and Hall resistivity ($\rho_{xx}$ and $\rho_{yx}$) are measured by a standard four-probe method using a variable temperature insert (VTI) with a superconducting magnet (Teslatron PT, Oxford Instruments) and commercial physical property measurement system (PPMS, Quantum Design). The samples are patterned into standard Hall bars for electrical transport measurements. The Hall resistivity $\rho_{yx}$ is defined as $V_y \cdot d/I_x$, where $I_x$, $V_y$, and $d$ stand for the current, transverse voltage and the thickness, respectively. The field dependence of the Hall resistivity is obtained after subtracting the longitudinal resistivity contribution, which presumably originates from slight misalignment of the Hall probes and is even-symmetric to the applied field.

The crystallinity of the film is examined by the XRD method. Figure 2(a) shows XRD patterns of the post-annealed Mn$_3$Sn (40 and 400 nm) films. Significantly, we confirm the films to be the single phase of Mn$_3$Sn and randomly oriented polycrystalline form; all the peaks can be indexed by the Bravais lattice with the lattice constants of $a$ = 5.67 Å and $c$ = 4.52 Å consistent with the hexagonal $P6_3/mmc$ symmetry of D0$_{19}$ Mn$_3$Sn, and there are no additional peaks coming from plausible impurity phases such as ferromagnetic Mn$_2$Sn and Mn$_3$Sn$_2$. The ratio of the peak intensity is found roughly consistent with the theoretical simulation (the bottom curve in Fig. 2(a)), except the ratio between (011) and (002) for the 400 nm film. This suggests that the 400 nm film has a slightly larger population of the grain having (011) plane parallel to the film surface than the ideal polycrystalline distribution. The surface morphology is examined by atomic force microscope (AFM) method. As can be seen in Fig. 2(b), the AFM scan image over 1 μm ×1 μm area of the post-annealed film (40 nm) reveals a relatively rough surface with a typical root mean square surface roughness of ~1.1 nm because of the multi-grain character.

To clarify the magnetic phase of the Mn$_3$Sn films and verify the absence of the ferromagnetic impurity as found in the XRD measurements, we perform the magnetization measurements. Figure 3(a) shows the temperature ($T$) dependence of the zero-field out-of-plane magnetization $M(H = 0)$ of the Mn$_3$Sn film obtained from 300 K to 100 K on cooling after the demagnetization process (we first applied a field of $\mu_0 H = 5$ T perpendicular to the film surface and decreased the field down to 0 T at 300 K). The zero-field magnetization exhibits an abrupt decrease below $T_1$ = 260 K, corresponding to the magnetic phase transition known for Mn$_3$Sn, namely, from the inverse triangular spin structure to the helical (spiral) spin structure on cooling [24,25]. This result provides strong evidence that the magnetic state is the same as the bulk Mn$_3$Sn and the large topological responses such as the anomalous Hall effect is expected in the high temperature phase with the inverse triangular spin structure. The spontaneous magnetization value at 100 K is as low as ~0.1 m$\mu_B$/Mn close to the resolution of our measurements. This further confirms the absence of the ferromagnetic impurity in our film. Figure 3(b) shows the field dependence of the magnetization obtained under field up to 5 T at room temperature. The magnetization $M$ exhibits a hysteresis with the spontaneous component of 6 m$\mu_B$/Mn (2.6kA/m) at $H = 0$. In the case of the single crystals of Mn$_3$Sn, the spontaneous $M$ is anisotropic; it is ~3 m$\mu_B$/Mn for the field along the (0001) plane (*ab*-plane), while it is absent for the [0001] direction (*c*-axis) [8]. Thus, the polycrystalline average is ~2 m$\mu_B$/Mn, which is in the order of our polycrystalline film.

Most significantly for the Weyl antiferromagnet Mn$_3$Sn, sizeable anomalous Hall effect emerges in the magnetic Weyl metal state with the inverse triangular spin structure. Figure 4(a) provides the $T$ dependence of the longitudinal resistivity $\rho_{xx}$ and the spontaneous Hall resistivity $\rho_{yx}(H = 0)$ under zero field of the Mn$_3$Sn thin film (40 nm). The resistivity $\rho_{xx}(T)$ confirms the metallic transport consistent with the previous work on the bulk single crystals [8,15-18]. The zero-field Hall resistivity $\rho_{yx}(H = 0)$ was obtained from the $(\rho_{yx}(H = +0) - \rho_{yx}(H = -0))/2$. Here we use +0 and −0 to indicate zero magnetic field approached respectively from the +5 T and −5 T in the measurement of the spontaneous component of the Hall resistivity at each temperature. Similar to the magnetization, we observe a sharp decrease of $\rho_{yx}(H = 0)$ below $T_1 = 260$ K on cooling; this agrees with the magnetic symmetry consideration in $Mn_3Sn$ that the inverse triangular spin structure breaks the global time-reversal symmetry while it is restored in the low temperature helical phase. Figure 4(b) indicates the field dependence of the Hall resistivity $\rho_{yx}$ at 300 K as a function of the magnetic field applied perpendicular to the film surface. The anomalous Hall effect of the 40 nm film exhibits a large change in $\rho_{yx}$ from 1.5 μΩcm to -1.5 μΩcm with increasing field because of the switching of the antiferromagnetic domain in the real space and the large fictitious field in the momentum space, and nearly the same field dependence is observed for the 400 nm film. The coercivity is found to be 0.6 T, consistent with the results seen in the magnetization curve (Fig. 3(b)). All the ferromagnetic impurities known for the Mn-Sn binary system has the Curie temperature lower than the room temperature. Thus, the large spontaneous Hall effect with the vanishingly small magnetization seen at 300 K and the $T$ dependence of the Hall effect and the magnetization having clear suppression below $T_1$ provide good evidence that the film is single phase and it is the inverse triangular spin state of the $Mn_3Sn$ polycrystalline phase at $T > T_1$ that exhibits the large anomalous Hall effect.

Here we note that no report has been made on the thin film of the antiferromagnetic $Mn_3X$ film exhibiting the large anomalous Hall effect, having the same order of magnitude as the value reported for bulk. In fact, the literature on the epitaxial thin film fabrication of $Mn_3Sn$ has been published [26], but without showing any results on anomalous Hall effect. They find that their film has one order larger magnetization than what has been seen in single crystals, suggesting the defect and/or inclusion of a ferromagnetic impurity. The comparison with our results indicates that the fabrication of the single phase, and appropriate Mn concentration are the keys for the observation of the anomalous Hall effect.

Our fabrication of the high-quality thin film of the Weyl antiferromagnet $Mn_3Sn$ and observation of its large anomalous Hall effect provide an important step for further developing devices useful for antiferromagnetic spintronics such as a high-speed and high-density information storage.

**Acknowledgments**

We thank T. Nishikawa, T. Ohtsuki, H. Man, Q. Ma, and S. Miwa for helpful discussions. We also thank R. Ishii for the ICP-AES measurements, and M. Lippmaa for the AFM measurements. This work is partially supported by CREST (JPMJCR15Q5), Japan Science and Technology Agency, by Grants-in-Aid for Scientific Research (16H02209), and by Grants-in-Aids for Scientific Research on Innovative Areas (15H05882, 15H05883 and 26103002) from the Ministry of Education, Culture, Sports, Science, and Technology of Japan, and Program for Advancing Strategic International Networks to Accelerate the Circulation of Talented Researchers (No. R2604) from the Japanese Society for the Promotion of Science (JSPS). D.Q. was supported by the Japan Society for the Promotion of Science (JSPS) as international research fellow. Work at JHU has been supported in part by SHINES Grant No. DE-SC0012670, an Energy Frontier Research Center from the U.S. Department of Energy, Office of Science, Basic Energy Science. The work at IQM was supported by the US Department of Energy, office of Basic Energy Sciences, Division of Materials Sciences and Engineering under grant DE-SC0019331.

**Figures**

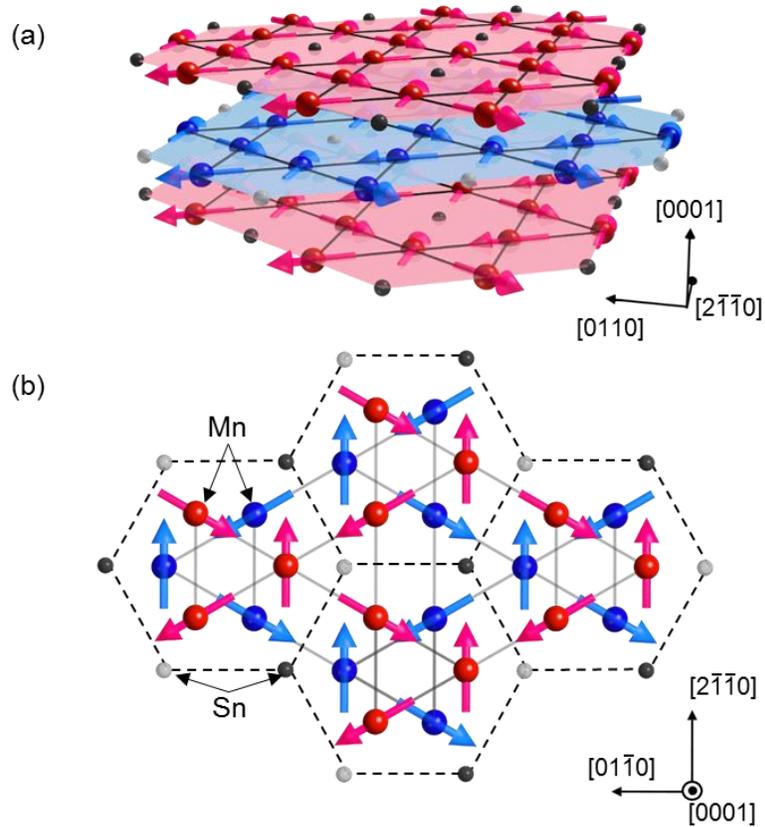

**Fig. 1.** 3D view (a) and top view along the (0001) direction (b) of crystal and magnetic structures of the Weyl antiferromagnet $Mn_3Sn$. The large red and blue spheres (small black and gray spheres) represent Mn atoms (Sn atoms) at $z = 0$ and $1/2$, respectively. The Mn magnetic moments (arrows) lie in the (0001)-plane and form an inverse triangular spin structure. The spin structure on the kagome bilayers can be considered as ferroic ordering of a cluster magnetic octupole. The octupole unit is represented by a hexagon in Fig. 1(b).

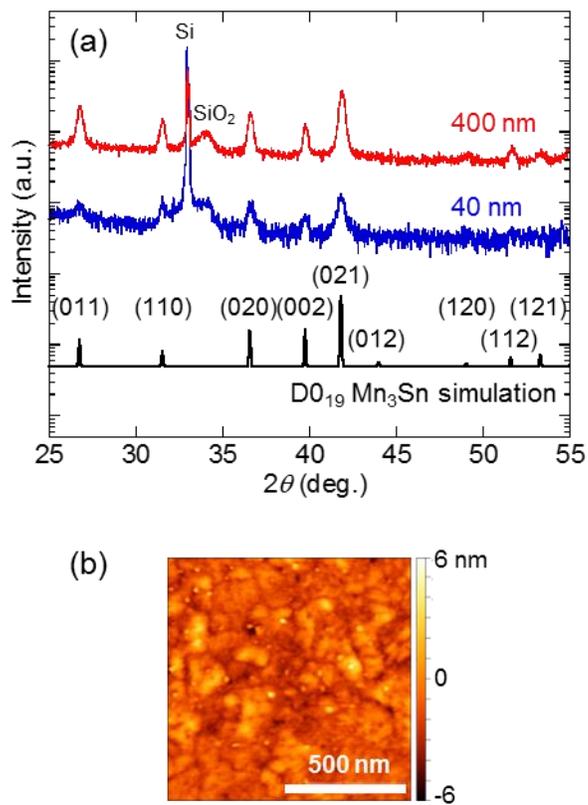

**Fig. 2.** (a) Room temperature spectra obtained by X-ray diffractometer (CuKα1 with a wave length of 1.54059 Å) for the 40 and 400 nm thin films of the $Mn_3Sn$ on a $Si/SiO_2$ substrate. The theoretical spectrum for $D0_{19}$ $Mn_3Sn$ structure is presented at the bottom. (b) Surface topography of the $Mn_3Sn$ (40 nm) thin film obtained by atomic force microscope (AFM). The scan area is 1μm ×1μm.

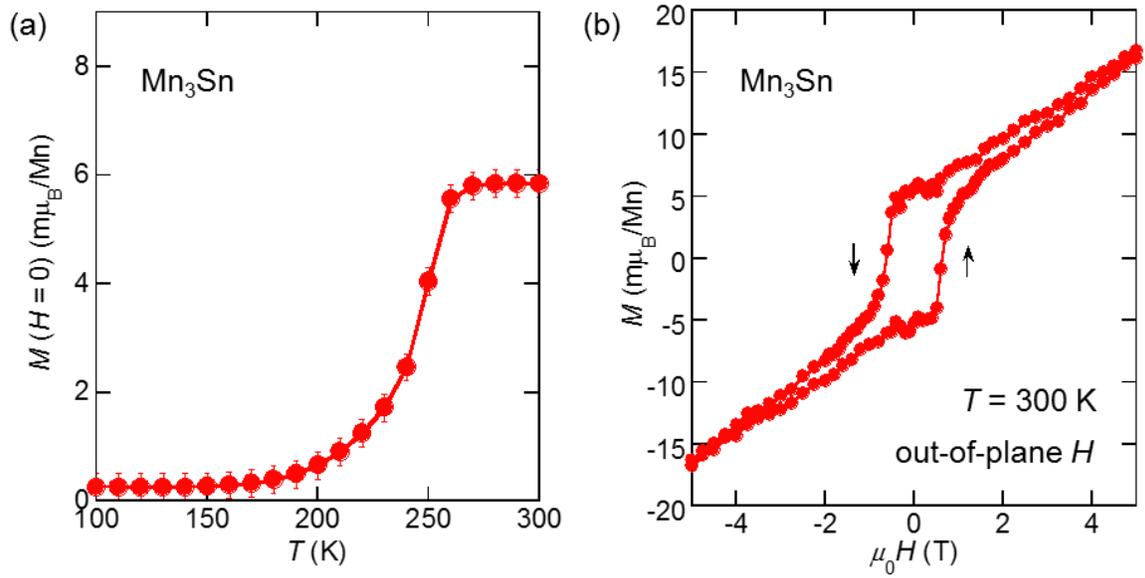

**Fig. 3.** (a) Temperature dependence of the zero-field perpendicular magnetization of the Mn$_3$Sn (400 nm) thin film. (b) Field dependence of the magnetization measured under the magnetic field applied in the perpendicular direction to the surface of the Mn$_3$Sn (400 nm) thin film. The arrows indicate the direction of the field sweep.

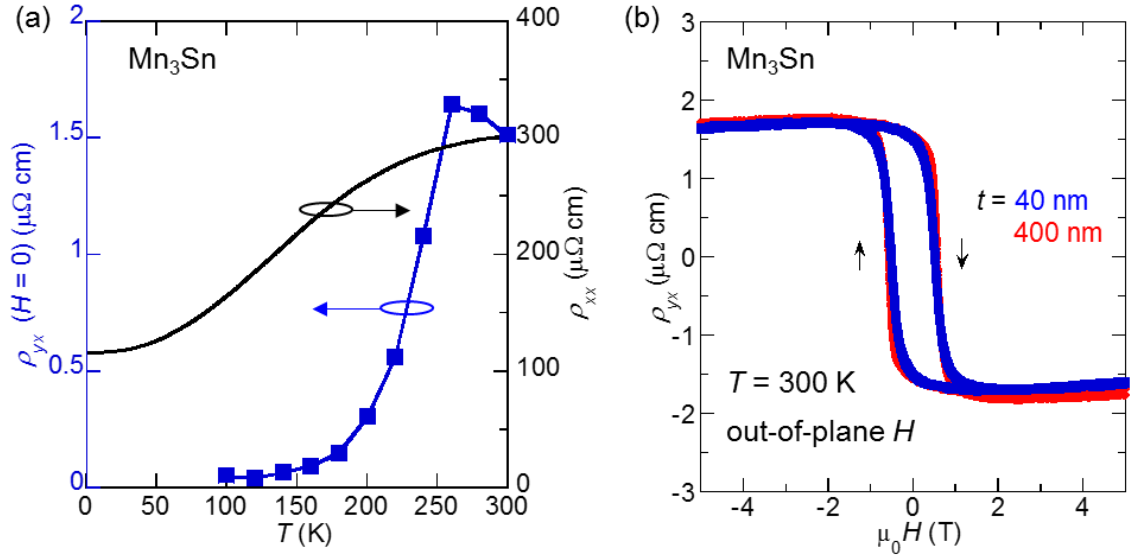

**Fig. 4.** (a) Temperature dependence of the resistivity and spontaneous Hall resistivity of the Mn$_3$Sn (40 nm) thin film. (b) Field dependence of the Hall resistivity $\rho_{yx}$ measured under the magnetic field applied in the perpendicular direction to the surface of the Mn$_3$Sn (40 and 400 nm) thin films. The sign of the Hall resistivity $\rho_{yx}$ is negative under a positive field larger than the coercivity. The arrows indicate the direction of the field sweep.